# Scheme for media conversion between electronic spin and photonic orbital angular momentum based on photonic nanocavity

**CHEE FAI FONG,**[1*] **YASUTOMO OTA,**[1] **SATOSHI IWAMOTO,**[1,2] **AND YASUHIKO ARAKAWA**[1]

[1] *Institute for Nano Quantum Information Electronics, The University of Tokyo, 4-6-1 Komaba, Meguro-ku, Tokyo 153-8505, Japan*
[2] *Institute of Industrial Science, The University of Tokyo, 4-6-1 Komaba, Meguro-ku, Tokyo 153-8505, Japan*
*\*cffong@iis.u-tokyo.ac.jp*

**Abstract:** Light with nonzero orbital angular momentum (OAM) or twisted light is promising for quantum communication applications such as OAM-entangled photonic qubits. There exist photonic OAM to photonic spin angular momentum (SAM), as well as photonic SAM to electronic SAM interfaces but not any direct photonic OAM-electronic SAM (flying to stationary) media converter within a single device. Here, we propose a scheme which converts photonic OAM to electronic SAM and vice versa within a single nanophotonic device. We employed a photonic crystal nanocavity with an embedded quantum dot (QD) which confines an electron spin as a stationary qubit. Spin polarized emission from the QD drive the rotation of the nanocavity modes via the strong optical spin-orbit interaction. The rotating modes then radiate light with nonzero OAM, allowing this device to serve as a transmitter. As this can be a unitary process, the time-reversed case enables the device to function as a receiver. This scheme could be generalized to other systems of resonator and quantum emitters such as a microdisk and defects in diamond for example. Our scheme shows the potential for realizing an (ultra)compact electronic SAM-photonic OAM interface to accommodate OAM as an additional degree of freedom for quantum information purposes.

## References and links

## 1. Introduction

Twisted light has nonzero orbital angular momentum (OAM) with a wave front characterized by $\exp(il\phi)$, indicating an azimuthal dependence in its phase, where $|l|$ denotes the order of OAM, also known as the topological charge, and $\phi$ is the azimuthal angle in the plane of the wave front. The wave front also has an intensity distribution with radial dependence which can be described using, for example, the Bessel or Laguerre-Gaussian modes. Twisted light has been studied extensively since its formal conception [1] with interest ranging from its fundamental properties [2–4], the role of OAM in the optical spin-orbit interaction [5] to the multitude of applications utilizing twisted light [6–9]. Garnering particular interest in the research community is the promise of OAM for optical communication, both in the classical [10] and quantum regime [11–14], due to the large availability of orthogonal OAM modes for spatial division multiplexing to increase data capacity.

Given the potential of flying photonic OAM to encode quantum information for long distance communication, in order to fulfill its potential as a new degree of freedom, it is imperative to interface such photonic OAM with a stationary counterpart such as a solid state spin qubit for local manipulation and storage purposes. To convert a photonic OAM to electronic spin angular momentum (SAM), with current technology, one would need two interfaces (Fig. 1(a)). The first interface converts the photonic OAM to photonic SAM. There are numerous reports of such interfaces which are usually optical elements made of glass polymer [15,16], liquid crystal [17], dielectric [18] or plasmonic metasurfaces [19,20]. After passing through the first interface, the output photon with SAM is then incident on a so-called photonic SAM-electronic SAM interface in which the key component is a quantum emitter, such as a quantum dot (QD) [21–23], an atom [24–26] or a nanoparticle [27–29]. The quantum emitter absorbs the photon, thereby converting the photon spin into electron spin. These two interfaces have to be used in combination to enable a two way conversion from flying twisted photons to stationary electronic spins.

In this paper, we propose a scheme which allows for an effectively "1-step" media conversion between flying photonic OAM to stationary electronic SAM and vice versa (Fig. 1(b)). Our scheme is based on a 2D air-holes H1 photonic crystal (PhC) nanocavity with an embedded QD. The time-reversible conversion between photonic OAM and electronic SAM is enabled by the underlying optical spin-orbit interaction (SOI) in the tightly confined rotating optical modes, together with the coupling between the nanocavity and the twisted light (photonic OAM) modes (Fig. 1(c)). The QD is our choice of quantum emitter here as it could preserve spin information, enabling high fidelity conversion between OAM and SAM. In this scheme, the twisted light drives the confined rotating modes in the PhC nanocavity. The propagating modes then give rise to photonic spin via spin-momentum locking [30] where the sign of the OAM is projected to a corresponding direction of the spin. The photonic spin then generates an electron spin within the coupled QD, completing the media conversion. In the opposite case, the emission of a QD with an electron spin drives the rotating modes, which in turn causes the nanocavity to radiate twisted light.

In the following sections, we will outline the principle of operation of our scheme. Also, we will describe how a single PhC nanocavity-QD device could function as both transmitter (SAM-to-OAM) and a receiver (OAM-to-SAM) under this scheme. In addition, we will include a discussion on the nanocavity Purcell effect and Q-factor for efficient QD-nanocavity mode coupling, as well as the role of the nanocavity farfield emission for nanocavity-photonic OAM mode coupling.

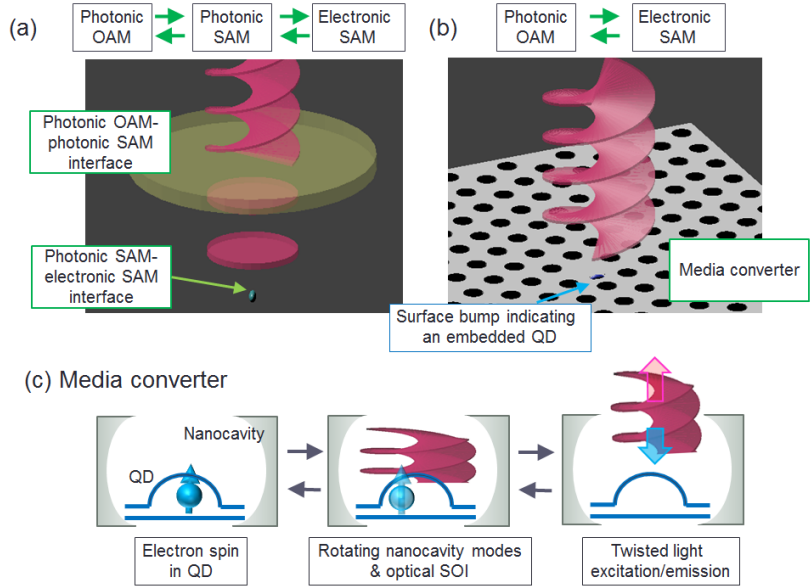

Fig. 1. (a) Schematic showing the conventional situation to convert photonic OAM to electronic SAM. Twisted light indicated by the helical phase (magenta) is incident on a photonic OAM-photonic SAM interface, converting the angular momentum, outputting a beam with constant phase fronts. This beam then excites a photonic SAM-electronic SAM interface, which is essentially a quantum emitter, to finally generate a stationary electron spin. (b) Schematic of the proposed media converter scheme which could convert photonic OAM to electronic SAM and vice versa with just a single nanophotonic device: a PhC nanocavity with an embedded QD. (c) Illustration of how the media conversion is enabled by the optical spin-orbit interaction within the nanocavity. Going from left to right, the light emission of a spin-polarized electron in a QD drives the nanocavity modes to rotate, resulting in the nanocavity emitting twisted light. In the opposite case, twisted light excitation drives the rotating modes, which then generates an electron spin in the QD.

## 2. Principle of operation of media conversion scheme

Our scheme is based on the H1 PhC nanocavity which consists of a missing air-hole in a triangular lattice of air-holes. The H1 nanocavity supports the degenerate quadrupole modes which are the modes of interest. A QD is placed within the nanocavity such that it couples with the quadrupole modes. We will formulate the stationary quadrupole modes, expressing them in terms of rotating wave components. Then, we will describe how suitable excitations can drive the stationary modes to rotate. The rotating modes will in turn give rise to locally circularly polarized electric fields, which can be quantified by calculating the SAM density. Based on the SAM density and the local field intensities, one can decide the suitable positions for the QD in order to achieve good media conversion fidelity and efficiency.

A 2D microdisk can be used to approximate the relevant H1 PhC nanocavity modes, retaining their most important features such as the mode profile, with significantly simplified mathematical expressions. The modes are expressed in terms of cylindrical harmonics to illustrate and explain the physical mechanisms for the media conversion between photonic OAM and electronic SAM. In this simplified 2D microdisk model, following from the Helmholtz equation $-\nabla^2 \psi = n^2 k^2 \psi$, solving in the cylindrical coordinates $(r,\phi,z)$ with appropriate azimuthal and radial boundary conditions gives wavefunctions in terms of cylindrical harmonics [31]

$$\psi_m(r,\phi) = a_m J_m(n,k,r) \exp(im\phi) \qquad (1)$$

where $J_m$ is the $m$th order Bessel function of the first kind and $m$ is the index of the angular momentum, $n$ being the refractive index, $k=\omega/c$ is the wavenumber and $a_m$ is the (complex) amplitude of the component such that $\psi_m$ is normalized. $\psi_m$ expresses a rotating wave where positive and negative values of $m$ correspond to clockwise and counter-clockwise rotating component respectively [32]. Here, we consider the TE polarization and that the microdisk is in the x-y plane, therefore $\psi_m$ represents the z-component of the magnetic field vector $H_z$.

We are interested in the quadrupole modes, $m = 2$ and these degenerate stationary modes can be expressed as a linear combinations of Eq. (1)

$$\begin{aligned} H_z^{Q1} &= \psi_{-2} + \psi_{+2} \\ H_z^{Q2} &= i(\psi_{-2} - \psi_{+2}) \end{aligned} \qquad (2)$$

where Q1 and Q2 denote the two eigenmodes. Eq. (2) are not normalized and $H_z^{Q2}$ is expressed with a phase factor $i$ such that both modes are real functions. These eigenmode equations satisfy the orthogonality condition $\int H_z^{Q1*} \cdot H_z^{Q2} dr^2$. The electric field can be determined from the magnetic field using $E(r) = \frac{i}{\omega \varepsilon_0 \varepsilon_r} \nabla \times H(r)$ where $\varepsilon_0$ and $\varepsilon_r$ are the vacuum permittivity and the relative permittivity respectively.

The quadrupole mode profiles obtained from the above analytic model is consistent with those obtained from the finite-difference time-domain (FDTD) numerical simulations of the H1 PhC nanocavity (Fig. 2(a)), indicating the validity of the analytical model. Simulations were performed using Synopsys' RSoft. In the simulations, we consider a number of key parameters of the air-holes [33] including the period, $a$ and radius $r$, as well as the modifications to the radius, $r_1'$ ($r_2'$) and the positions, $d1$ ($d2$) of the first (second) nearest air-holes (Fig. 2(b)). The FDTD simulations of H1 PhC were performed with the following parameters: $r/a = 0.289$, $r_1' = 0.25a$, $d1 = 0.15a$, $r_2' = 0.30a$ and $d2 = -0.08$, with $a = 0.3\mu m$. The slab is made out of GaAs with refractive index $n = 3.4$ and the slab thickness is $0.366a$, giving a normalized frequency of the modes ($a/\lambda_{cav}$) at 0.278 ($\lambda_{cav} \sim 927nm$) with a Q-factor of about 1800 and a mode volume $V_{mode} \sim 0.6(\lambda_{cav}/n)^3$. Simulations were performed with a grid size of $a/20$. The effect of the reduction of the first nearest holes radius and moving them outwards away from the center is twofold: increase in the quadrupole emission wavelength and also the Q-factor. The modifications to the second nearest air-holes are mainly to tune the Q-factor, with a small effect on the mode emission wavelength.

The quadrupole modes have field distributions which in combination could give rise to radiation with a donut-like intensity distribution, as well as azimuthally varying phase profiles in its wavefront, which are characteristics of twisted light. Indeed, we have found that the coupling between the PhC nanocavity quadrupole modes and twisted light excitation is allowed by the conservation of total angular momentum, therefore the quadrupole modes are suitable for this scheme (see Appendix B for further details).

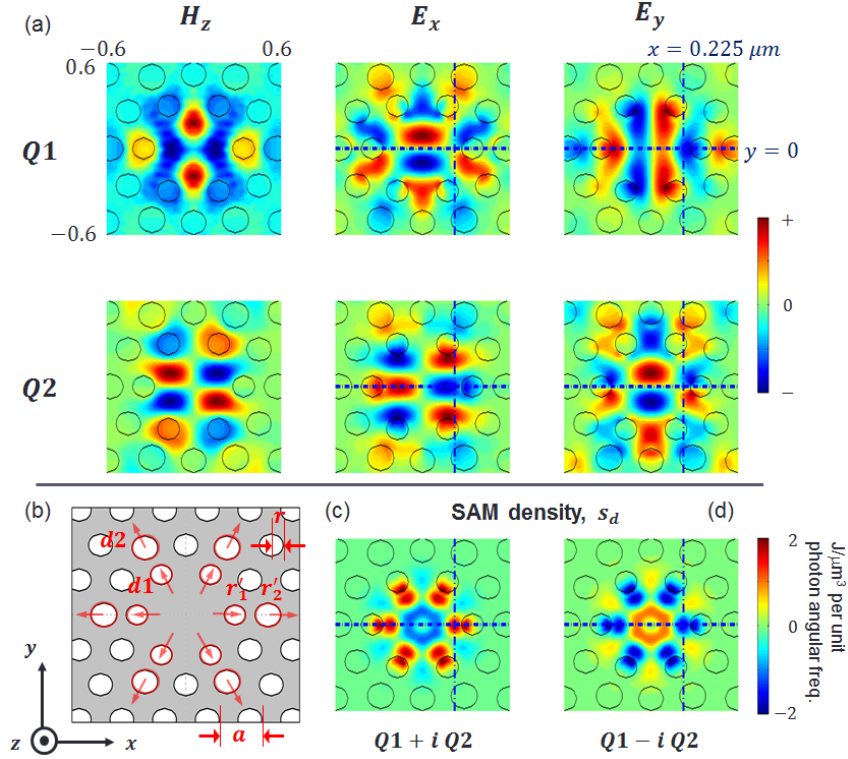

Fig. 2. (a) Quadrupole mode profiles of the relevant field components – $H_z$, $E_x$ and $E_y$ – at the center of the PhC slab (y = 0) in the TE polarization obtained from FDTD numerical simulations. The black circles outline the edges of the air-holes of the PhC. The software RSoft was used for all FDTD simulations. The dashed lines mark along x = 0.225 and y = 0 where the intersection serves as the exemplary position for the discussions in the main text. (b) Schematic of PhC nanocavity. Labelled are the key parameters of the air-holes. See main text for parameter description and values. The SAM density distribution at the PhC slab center for the case of (c) Q1 + $i$Q2 and (d) Q1 − $i$Q2 calculated using the field components in (a). The SAM density takes the unit of energy density [J/μm$^3$] per unit photon angular frequency.

By driving the modes – with an circular dipole within the modes (transmitter case) or external twisted light excitation (receiver case) – under the right conditions the two quadrupole modes can be excited with a relative phase difference of ±π/2. Spin polarized emission of a QD can play the role of a circular dipole where the electric field components have a relative phase difference: $E_x \pm iE_y$ corresponding to right and left circular dipole respectively [21]. By embedding such a circular dipole in the nanocavity, it could excite the two quadrupole modes with a corresponding ±π/2 phase difference. On the other hand, twisted light excitation only couples to one of the rotating wave component at a time due to the conservation of angular momentum. Therefore, according to our formulation of the modes in Eq. (2), the twisted light excites either the $\psi_{+2}$ or $\psi_{-2}$ rotating wave component of both Q1 and Q2, giving rise to the ±π/2 phase difference. For both cases, the excitations give resulting modes which are linear combinations of the quadrupole modes Q1± $i$Q2, where the fields can be written as

$$H_z^{\pm} = \frac{1}{2}\left(H_z^{Q1} \pm i H_z^{Q2}\right) = \psi_{\pm 2} \qquad (3)$$

including the factor of 1/2 for normalization. The electric field components $E^{\pm}$ can be expressed in similar linear combinations. The resulting modes are no longer standing waves but rotating waves. The equal excitation of the two quadrupole modes with ±π/2 phase difference drives the modes to rotate (anti)clockwise as indicated by $\psi_{\pm 2}$. Indeed, the rotation of such linear combination of the quadrupole modes is observed in the FDTD numerical simulations, illustrated in Fig. 3, showing the (anti)clockwise rotation as expected.

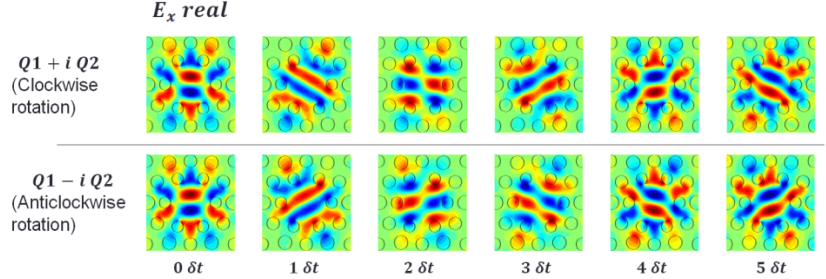

Fig. 3. The real part of $E_x$ of the linear combination of quadrupole modes Q1 ± $i$Q2 at different time steps, $\delta t$ is plotted illustrating the expected rotation as a labelled. Similar behaviour is observed in other field components (not shown). $\delta t$ corresponds to the time it takes for light to propagate $c\delta t$~0.1μm, about 30 $fs$ in absolute time, which is 1/8$^{th}$ of a rotation period in the simulations here.

Due to the tight confinement in the nanocavity, the polarization (spin) and the spatial distribution of the fields are no longer independently conserved, representing significant optical SOI. Consequently, the rotating modes give rise to locally rotating electric fields which can be thought of as chiral points or out-of-plane (transverse) spin. The amount of local circular polarization of the fields is related to the SAM density [34] given by the following equation:

$$s_d(r) = \text{Im}[\varepsilon_0 \varepsilon_r E^* \times E + \mu_0 \mu_r H^* \times H]/4\omega \qquad (4)$$

where $\varepsilon_0$ ($\varepsilon_r$) and $\mu_0$ ($\mu_r$) are the vacuum (relative) permittivity and vacuum (relative) permeability respectively. Here in a nanocavity, the $s_d$ can be considered as the difference of the field intensities of right and left circular polarization. Figure 2 (c, d) shows the distribution of $s_d$ at the slab center (y = 0) within the PhC nanocavity calculated using the linear combinations of the fields (Eq. (3)). The $s_d$ distribution reflects the C6v symmetry in accordance to the symmetry of the PhC nanocavity. The regions with the largest $s_d$ are distributed close to the first nearest air-holes. In the TE polarization, the relevant fields of $H_z$, $E_x$ and $E_y$ naturally result in $s_d$ pointing in the out of plane ±z-direction. The direction of the spin as indicated by the $s_d$ distributions for the two coupled modes are opposite of each other as expected, reflecting the spin-momentum locking. In other words, the direction of mode rotation defines the direction of the transverse spin.

To facilitate the media conversion to and from an electron spin, the QD needs to be placed at a suitable position within the PhC nanocavity. The conversion fidelity – how well the degree of photonic OAM is being converted to electronic SAM and vice versa – is related to the local degree of circular polarization of the fields within the nanocavity. At the position of the QD, the closer the local DOP is to |1| to better the conversion fidelity will be. The local DOP can be obtained by calculating $s_d/W$, where $W$ is the time-averaged total field energy density per unit angular frequency given by $(\varepsilon_0 \varepsilon_r E^* \cdot E + \mu_0 \mu_r H^* \cdot H)/4\omega$ [35,36]. The calculated DOP take values between ±1 (see Fig. 6 in Sec. 4) where ±1 correspond to right and left circular polarization respectively. In addition to the conversion fidelity, one should also consider the conversion efficiency which is related to the rate of emission of twisted photon or the rate of generation of electron spin in their respective media conversion cases. As will be discussed in further details in the next section, the conversion efficiency is partly governed by the electric field intensity or energy density within the nanocavity. The position with higher field intensity is favorable for better efficiency. Due to the rotating nature of the modes, the resulting $s_d$, DOP and electric field intensity all take similar C6v distribution. As such that the positions with the maximum $s_d$, DOP and electric field intensity are found to practically coincide. The positions where the maximum $s_d$, DOP and electric field intensity coincides are suitable positions to place a QD. The spatial coupling of the QD and the nanocavity can be realized via a number of methods including pre- [34] and post-selection of QD [32] relative to PhC fabrication.

In this proposed scheme, the presence of rotating modes which in turn give rise to transverse spin, together with a coupled quantum emitter are the key requirements for angular momentum media conversion. In the following sections, we will describe the further details of how a H1 PhC nanocavity with embedded QD media converter device can operate as a twisted light transmitter and receiver.

## 3. Twisted light transmitter: electronic SAM to photonic OAM conversion

For the twisted light transmitter case, the electronic SAM within a QD in the PhC nanocavity is effectively converted into photonic OAM in the emission. In order to achieve this, light emission from a QD with a confined electron spin plays the role of the circular dipole. The QD needs to be placed at a suitable position within the nanocavity to drive the rotation of the modes. Then, the nanocavity will radiate light with a helical phase, indicating successful media conversion of electronic SAM to photonic OAM.

Here we approximate the circular dipole as a point source which is valid in the case of a QD [37]. This will ensure that the optical selection rules are obeyed, such that each direction of the electron spin and its associated circular polarization emission is maintained [38]. For our PhC nanocavity device with the parameters as listed in Sec. 2, an example of a suitable position is at (x, y) = (0.225, 0) where the DOP is the maximum at |0.97|. At this position, a QD could excite the two degenerate quadrupole modes with a $\pi/2$ phase difference. This can also be inferred by looking at the field components at this position, for example, here the $E_x^{Q1}$ has zero amplitude while $E_x^{Q2}$ has significant nonzero amplitude, with $E_y^{Q1}$ and $E_y^{Q2}$ showing the opposite tendency of nonzero amplitude and zero amplitude respectively. Therefore, a circular dipole at (0.225, 0) could drive the quadrupole modes with the corresponding $\pi/2$ phase difference.

Figure 4 shows the nanocavity emission properties under such circular dipole excitation. Looking at the $E_x$ field component in a plane at a distance which is about twice the cavity resonant wavelength ~$2\lambda_{cav}$, the real part indicates that the field is rotating. More importantly, the phase at the center of emission varies from 0 to $2\pi$ azimuthally (anti)clockwise under right (left) circular dipole excitation. Furthermore, the farfield patterns of the emission show a donut-like profile reminiscent of twisted light. Excitation with a dipole of linear polarization does not rotate the modes and thus the emission does not show the azimuthally varying phase.

Both the conversion fidelity and efficiency can be discussed in two parts: 1) between the QD and the quadrupole modes and 2) between the quadrupole modes and photonic OAM mode. We will first discuss the conversion fidelity. In a QD, the confined spin down (up) corresponds to a right (left) circular dipole due to its optical selection rules. Therefore, experimentally, the helicity of the circular dipole within the nanocavity can be controlled via the helicity of the optical excitation of the QD. By placing the QD at (0.225, 0), with the QD emission in resonance with the cavity emission, the helicity of the excitation will determine the sign of the transverse spin at that position, as well as the sign of the OAM and the circular polarization of the emitted light. A perfect circular dipole – the light emission of a perfectly polarized electron in the QD – will rotate the modes and generate a transverse spin with fidelity of 0.97 at the said position as determined by the DOP.

The rotating quadrupole modes then give rise to photonic OAM in the emission. To determine the conversion fidelity between the quadrupole modes and photonic OAM, one needs to quantify the OAM in the emission. To do so, one can start by making use of the OAM density which is expressed as [4,34]:

$$l_d = r \times p_0 \tag{5}$$

with the orbital linear momentum density $p_0$ being

$$p_0 = \text{Im}[\varepsilon_0 \varepsilon_r E^* \cdot \nabla E + \mu_0 \mu_r H^* \cdot \nabla H]/4\omega \qquad (6)$$

Here we are concerned with the intrinsic OAM, where we take the origin of $r$ to be at (x, y) = (0, 0). The mean OAM content in the z-direction can then be obtained from:

$$l = \frac{\iint l_d \cdot \hat{z} \, r dr d\varphi}{\iint W \, r dr d\varphi} \qquad (7)$$

For the numerical simulations in Fig. 4, the mean OAM of the emission obtained under right (left) circular dipole excitation at (0.225, 0) is $l = 1.09$ (-1.04) respectively confirming the emission of twisted light. The helicity of the circular dipole therefore determines the resulting order of the OAM of the emission, in accordance to the optical SOI. The mean SAM content, $s$ or degree of circular polarization (DOP) of the emission can similarly be obtained with an equation analogous to Eq. (7), replacing $l_d$ with $s_d$. In the emission, $s = 0.81$ (-0.77) for the right (left) circular dipole excitation respectively. This shows that the emission is largely co-polarized with the excitation. The non-integer values of $l$ and $s$ is most likely due to the optical SOI where neither quantity is conserved independently. Instead, one could look at the total angular momentum of the emission $j = l + s$ where we have found it to be approximately ±2 for the respective cases, as predicted from the conservation of total angular momentum (see Appendix B). Overall, the conversion from an electron spin to photonic OAM has fidelity of close to unity.

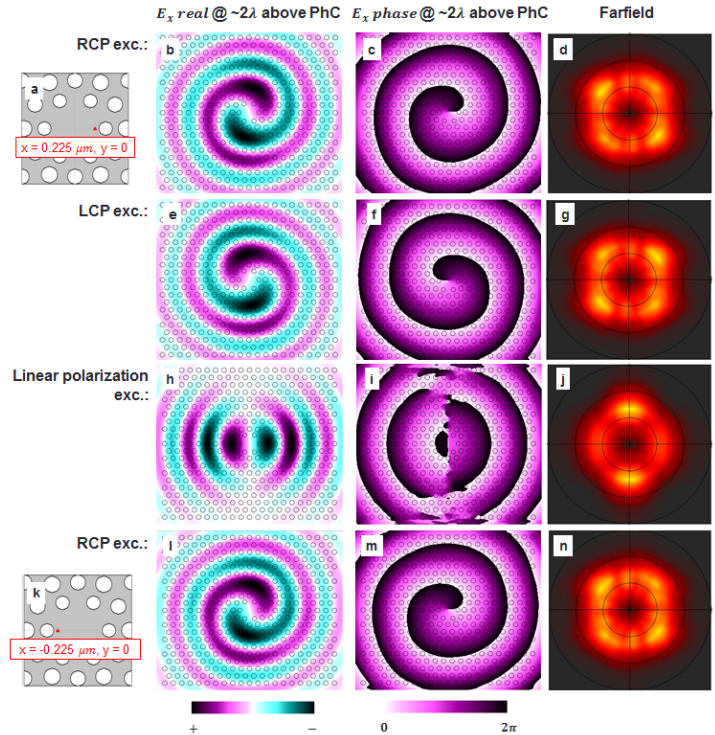

Fig. 4. (a) Schematic of the H1 PhC nanocavity showing the position of the QD at (x, y) = (0.225, 0). Plot of the real part of $E_x$, phase component of $E_x$ in the plane which is at ~$2\lambda_{cav}$ above the PhC, as well as the farfield under RCP excitation (b-d) and LCP excitation (e-g), linear polarization excitation (h-j) respectively, as well as the case of RCP excitation at (-0.225, 0) which is another suitable position for a QD (k-n) showing consistent field emission properties as that with excitation at (0.225, 0).

Similar to the conversion fidelity, for the conversion efficiency we first address the coupling between the QD and the quadrupole modes. The PhC nanocavity serves to enhance the coupling between the QD and the cavity modes while reducing coupling to other leaky modes. This enhancement is encapsulated by the Purcell factor, $F_p = \frac{3}{4\pi^2}\left(\frac{\lambda_{cav}}{n}\right)^3 \frac{Q}{V_{mode}}$ [39], as well as by the photonic bandgap effect [40]. For the parameters of the PhC nanocavity used here (Sec. 2), $Q\sim 1800$, $V_{mode}\sim 0.6(\lambda_{cav}/n)^3$ and thus giving $F_p\sim 220$. Aside from $F_p$, to evaluate the coupling enhancement, one also needs to factor in the local electric field intensity, $|E|^2_{local} = |E(r)|^2/|E|^2_{max}$ in consideration of the local density of states. For a QD at (0.225, 0), we found $|E(r)|^2/|E|^2_{max}\sim 0.8$, which is in fact the highest intensity within the nanocavity ($|E|^2_{max}$ lies just beyond the edge of the first nearest air-holes and is thus inaccessible). The $F_p$ considered here is usually applied to stationary modes where the field distribution is time independent, as opposed to the rotating modes. Nonetheless, the rotating modes here can be expressed in terms of stationary modes Eq. (3) and thus $|E|^2_{local}$ should be independent of time. This is supported by our simulation results where $|E|^2_{local}$ at various time steps is practically constant, showing minimal fluctuation of the order of 0.01. Moreover, our scheme here works in the single photon and single mode regime which further validates our choice of $F_p$. Considering both quantities, we found the coupling enhancement to be $F_p|E|^2_{local} \sim 170$. The coupling efficiency of the QD to the quadrupole modes can be written as $\beta = \gamma_{cav}/(\gamma_{cav}+\gamma_{leak})$ where $\gamma_{cav(leak)}$ represents the QD spontaneous emission rate into the cavity (leaky) modes. Given that $\gamma_{cav} = F_p\gamma_{bulk}$ and $\gamma_{leak}\sim 0.1\gamma_{bulk}$ [39], assuming perfect spectral coupling between the QD and the quadrupole modes, the coupling efficiency can be expressed as $\beta = F_p|E|^2_{local}/(F_p|E|^2_{local}+0.1)$. Therefore, our device here with is expected to be able to achieve a high coupling efficiency of $\beta\sim 1$ within the nanocavity. The modification to the air-holes around the nanocavity could be used to increase the Q-factor and thus improve the QD-nanocavity coupling efficiency.

As for the coupling between the quadrupole modes and photonic OAM mode, we limit our discussion here to the coupling between the mode emission and Laguerre-Gaussian OAM mode laser beam (see Appendix A). One can consider the overlap integral between the normalized twisted light field distribution, $E^{TL}$ and the farfield pattern of the mode emission $E^{ff}$, as such $\left|\int E^{ff*}E^{TL} rdrd\phi\right|^2$. The farfield emission is assumed to be collected and collimated with a lens – projecting the magnitude and the phase of the field components – such that the overlap integral with the twisted light excitation field can be calculated in the cylindrical coordinates. For the design parameters used here, the coupling efficiency reaches to about 30% for an OAM beam with optimized beam diameter of 1.3μm. The closer the farfield emission pattern is to that of the donut-like profile of the twisted light excitation, the better we can expect the coupling with the photonic OAM mode to be. The nanocavity can be designed in such a way that the farfield emission pattern takes a donut-like profile as much as possible, which will be left for future work.

Here we consider a symmetric PhC slab about the plane at the center of the slab, therefore emission in ±z will be symmetric as well. In light of this, in order to increase the twisted light emission efficiency, one could modify the PhC nanocavity by adding periodic shallow holes on the slab surface which functions as a second-order grating for directional emission [41]. Alternatively, one can consider adding bottom distributed back reflectors to redirect the downwards emission upwards and thus improving emission efficiency [42].

In a fabricated PhC nanocavity, the quadrupole modes could sometimes be split and are thus not degenerate due to fabrication error, although the degeneracy of the quadrupole modes could possibility be restored post-fabrication [35,36]. In the case of split modes, the QD emission should be such that it overlaps with both quadrupole modes spectrally in order to excite them simultaneously but at the expense of reduced coupling to the modes. We have verified this via simulation: we introduced a slight asymmetry in the PhC nanocavity to induce a splitting of ~0.5nm between the quadrupole modes. At $\lambda_{cav}$~927nm with $Q = \lambda_{cav}/\Delta\lambda_{FWHM}$ ~1800, the modes should have full widths at half maximum $\Delta\lambda_{FWHM}$~0.5nm and thus should be overlapped with each other. By setting our circular dipole to be at the center wavelength of the two quadrupole modes, the emission is indeed twisted light (not shown), without major degradation in the conversion fidelity. The spectral overlap of the QD and nanocavity modes can be ensured via the design of the PhC, as well as the tuning of either the emission wavelength of the QD [37] or the nanocavity modes [38].

## 4. Twisted light receiver: photonic OAM to electronic SAM conversion

The twisted light receiver case can largely be considered as the time-reversed of the transmitter case, converting photonic OAM to electronic SAM. Following from the time reversibility of the media conversion, a pair of the PhC nanocavity media converter of the same design can be interfaced with each other, where one functions as a transmitter and the other as a receiver. In this case, one can expect ideal coupling efficiency between the farfield radiation of the transmitter and the quadrupole modes of the receiver and therefore achieving both conversion efficiency and fidelity approaching unity. The focus of this section is to instead discuss a more general case where we consider the coupling of the nanocavity quadrupole modes to conventional Bessel or Laguerre-Gaussian shaped OAM beam.

Specifically, we consider the situation where twisted light of $l = \pm 1$ excites the center of the nanocavity at normal incidence from above. In other words, the center of wavefront of an external twisted light excitation coincides with the center of the PhC nanocavity (Fig. 5(g)). Custom data files containing the amplitude and the phase information were generated for use in the 3D FDTD simulations of twisted light excitation on a PhC cavity (see Appendix A for further details). The mean OAM (SAM) of the excitation calculated using Eq. (7) is confirmed to be $\pm 0.998$ and $\sim 10^{-7}$ for the nonzero (circularly polarized) and zero OAM (linearly polarized) cases respectively. In the simulations the beam spot size is set to be about 1μm consistent with actual laser spot sizes after focusing with an objective lens. We limit our discussion here to centered beams as off-centered beams could deteriorate the coupling with the quadrupole modes since they can be considered to consist of a superposition of centered beams with different values of $l$ [43].

The twisted light excitation couples to the constituent rotating wave components of the quadrupole modes due to the conservation of total angular momentum which cause the modes to rotate as described in Sec. 2. By looking at the mode profiles at different time steps (figure not shown), we found that the modes are indeed rotating and that the modes co-propagate in accordance to the sign of OAM. Twisted light of $l = +1(-1)$ excites (anti) clockwise rotating modes, which once again give rise to transverse spin.

Figure 5 shows the $s_d$ distribution within the nanocavity under twisted light excitation of right circular polarization (RCP, $s = +1$), left circular polarization (LCP, $s = -1$) and linear polarization ($s = 0$). The excitation of $l = +1$, RCP (Fig. 5(a)) and $l = -1$, LCP (Fig. 5(e)) are the well-coupled cases which gives the largest magnitudes of $s_d$, with distributions of opposite signs consistent with the opposite signs of $l$. For the cases of $l = +1$, LCP (Fig. 5(b)) and $l = -1$, RCP (Fig. 5(d)), there is no significant $s_d$ indicating that these excitation light do not couple to the quadrupole modes. Note that the well-coupled cases have $j = \pm 2$ as opposed to $j = 0$ of the non-coupling cases. Linearly polarized twisted light excitations with $j = \pm 1$ (Fig. 5(c, f)) couples to the quadrupole modes: while the $s_d$ has essentially the same distribution, the magnitude is 1/2 of that of their respective well-coupled counterparts. These observations can be explained in line with the conversation of the total angular momentum. For the case of $j = \pm 2$, all of the twisted light could couple to the quadrupole modes (angular momentum index $m = 2$) and thus giving the largest values of $s_d$. Twisted light of $j = \pm 1$ still couple to the quadrupole modes as linear polarizations can be thought of to be composed of constituents

RCP and LCP. Under the same excitation intensity, in each case of $j = \pm1$, only half of the twisted light field energy couples with the quadrupole modes which in turn results in the 1/2 factor in the magnitude of $s_d$. As for the $j = 0$ cases, the excitation does not couple with the quadrupole modes meaning such coupling does not conserve $j$. From these observations, we can deduce that the allowed exchange of angular momentum $\Delta j$ between the excitation and the quadrupole modes are $0, \pm1$. Given how the conservation of total angular momentum governs the coupling between the twisted light and the nanocavity modes, the conservation law can be used to make predictions or to aid choices of the type of excitation and modes to use in such media converter scheme even in other cavity-emitter systems.

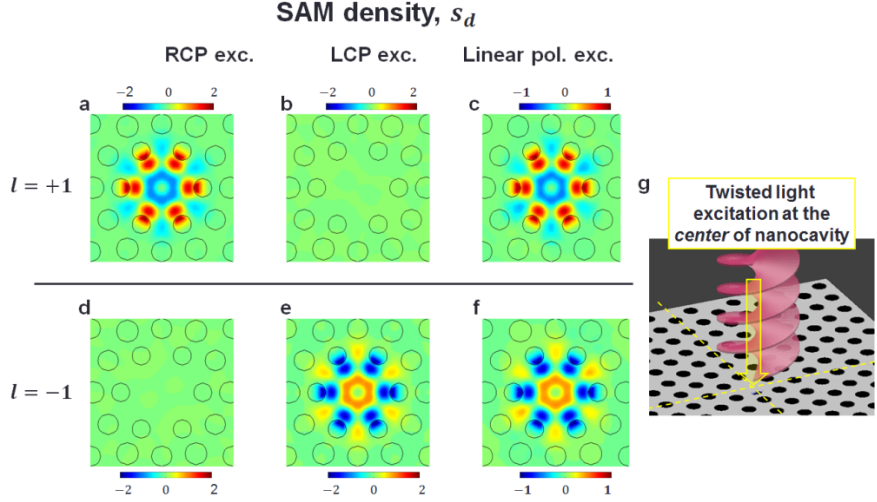

Fig. 5. Plots of the normalized SAM density of the quadrupole modes at the center of the PhC slab for the case of twisted light excitation with $l = +1$(a-e) and $l = -1$ (f-j) with different polarization as labelled. Note the scale of the color bar. (g) Schematic illustration showing that the excitation configuration considered here where an external twisted light excitation from above the PhC is at normal incidence at the center of the nanocavity.

Similarly, the local DOP within the nanocavity can be calculated using $s_d/W$ (Fig. 6). Both circular and linear polarization excitations give similar distribution of local DOP. Despite the lower magnitude of $s_d$ due to reduced coupling under linear polarization excitation, the magnitude of the total field energy flux densities varies accordingly to give DOP values between $\pm1$. As mentioned in Sec. 2, the positions with largest $|s_d|$ and the largest DOP magnitude tend to coincide and such positions are suitable to place a QD.

With the point dipole approximation for a QD, the local value of DOP can be taken as a measure of the conversion fidelity which signifies how well-polarized the generated electron spin can be in the QD. Following from this, based on our exemplary numerical simulation results and placing a QD at (0.225, 0), the twisted light excitation could generate spin-polarized electrons within the QD with a high fidelity of 0.97. Due to the optical SOI, the local chiral points in the nanocavity are expected to be completely circularly polarized. The maximum local DOP not reaching reach $|1|$ could be a numerical error due to the grid nature of the simulation. The local DOP of $|1|$ could be at a neighboring position which is inaccessible given the current grid size resolution ($a/20$).

For this receiver case, the conversion efficiency can be discussed in terms of 1) coupling between the twisted light excitation with the quadrupole modes and 2) generation of the electron spin in QD. Given the time-reversible nature of the media conversion, the coupling strength between the excitation and the quadrupole modes can be addressed as in the transmitter case: by comparing how well the twisted light field distribution matches the farfield distribution of the quadrupole modes emission; the better the match, the better the coupling. The coupling efficiency is expected to be ~30% for the well-coupled $j = \pm 2$ cases and half of that for $j = \pm 1$ cases. Therefore, there is further room for the optimization of the farfield emission.

The efficiency of the generation of the electron spin in QD will similarly depend on the Purcell effect. The $F_P |E|^2_{local}$ which we have found to be about 170 for our nanocavity design implies the significant enhancement of the density of states of cavity mode photon which could couple to the QD electronic states, which could in turn allows for a high coupling efficiency of $\beta \sim 1$ within the nanocavity. The high $\beta$, together with the regions with large electric field intensity promises a good enhancement of the rate of generation of electron spin due to the efficient coupling between the QD and quadrupole modes.

Regions within the nanocavity with large electric field intensity and local DOP will therefore give efficient media conversion of angular momentum with high fidelity. The generated spin can then be readout via, for example, coupled waveguides [44,45] or tunneling electron current [46,47].

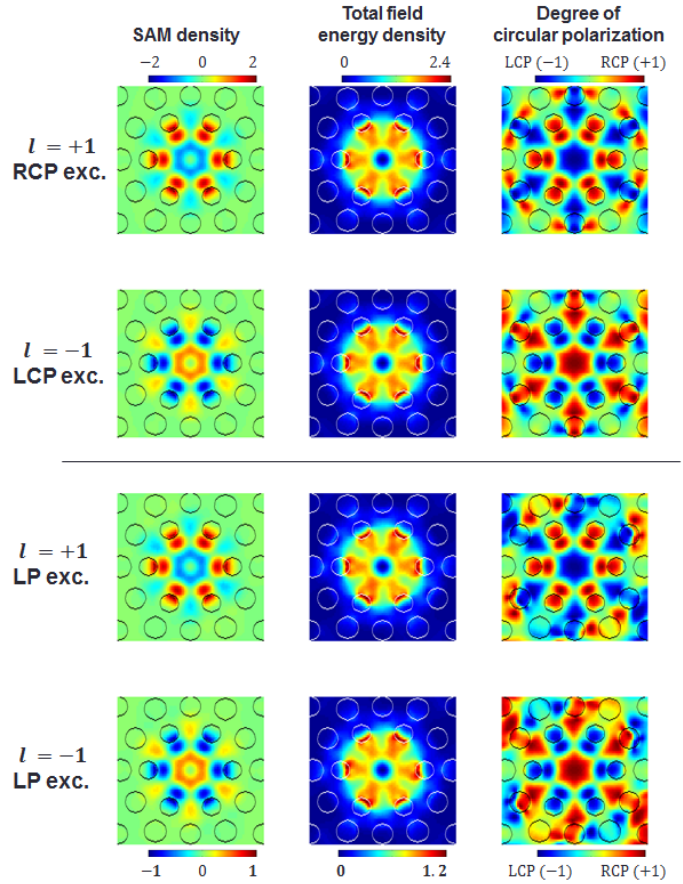

Fig. 6. SAM density, total field energy density and the degree of circular polarization of under excitation of $l = \pm 1$, together with the polarization as labelled.

## 5. Summary

In summary, we have described a scheme for media conversion between stationary electronic SAM and flying photonic OAM based on a PhC nanocavity, enabling an effectively "1-step" media conversion with a single nanophotonic device as opposed to using two conversion interfaces which often involves bulky optics. The key requirements are the rotation of the quadrupole modes which in turn give rise to locally circularly polarized field and a coupled QD which facilitates the confinement or generation of an electron spin. The direction of the rotation of the modes will determine the direction of the transverse spin, an example of spin-momentum locking due to the underlying optical SOI. We have described the H1 PhC nanocavity with embedded QD media converter device and how it could function as both a twisted light transmitter and a receiver. In the transmitter case, the spin polarized emission of the QD acts as the circular dipole which drives the rotation of the quadrupole modes. The resulting cavity emission has nonzero OAM. In the receiver scheme, twisted light excitation drives the rotation of the quadrupole modes, allowing for the generation of electron spin within a coupled QD.

The scheme promises good conversion fidelity of close to unity due to the small mode volume of a nanocavity which ensures strong optical SOI. Furthermore, this scheme also shows the potential for good conversion efficiency: even with a moderately large Q-factor, we can expect a high coupling efficiency between the quadrupole modes and the circular dipole. For the specific design of the PhC nanocavity here, we obtained a coupling efficiency of ~30% between the quadrupole modes and the external photonic OAM mode $e.g.$ twisted light excitation. It is worth emphasizing that both the fidelity and efficiency can be further optimized by improving the design of the PhC nanocavity. Given the versatility of this scheme, a more detailed formulation of the conversion fidelity and efficiency, as well as further optimization is worthy of future exploration.

While we have presented details of a device using H1 PhC nanocavity and QD, this scheme can be generalized to work in other types of system such as a microdisk cavity, together with other quantum emitters including defect states in SiC or diamond, atoms and so on. This scheme would be useful for quantum communication; in particular this scheme could function as a spin-photon-OAM interface to enable the utilization of OAM as a new degree of freedom for encoding information to increase data capacity. Such PhC nanocavity media converters working in pairs or more with ideal coupling to each other could also realize quantum repeaters [48] or be used for coupling to quantum memories. The compatibility of this scheme with conventional semiconductor growth and fabrication technologies also makes it attractive for use in integrated optical circuits.

## Appendix A: Twisted light excitation source

Taking the z-direction to be the direction of propagation, the vector potential of the twisted light in the Coulomb gauge is given by [49]:

$$A(r,t) \simeq \varepsilon_\sigma A_0 J_l(k_r r) \exp(il\phi) \exp[i(k_z z - \omega t)] \tag{8}$$

where $\varepsilon_\sigma$ is the polarization vector, $A_0$ is the (complex) amplitude factor, $J_l$ is the $l$th order of the Bessel function of the first kind where $l$ is the topological charge, while $k_z$ and $k_r$ are the z- and transverse wavevector respectively. The twisted light beams carry an OAM of $l\hbar$ per photon [1]. The circular polarization vector is expressed as $\varepsilon_\pm = \hat{x} \pm i\hat{y} = e^{\pm i\phi}(\hat{r} \pm i\hat{\phi})$ in the Cartesian and cylindrical coordinate basis. Here we consider a beam in the paraxial regime such that $k_z \gg k_r$, therefore the OAM and the polarization can be treated separately. Twisted light has a phase singularity at the center of the beam spot and thus zero intensity (Fig. 7). The electric field of the twisted light is then

$$E^{TL}(r,t) = -\frac{\partial}{\partial t} A(r,t) = -i\omega \varepsilon_\sigma A_0 J_l(k_r r) \exp(il\phi) \exp[i(k_z z - \omega t)].$$

## Appendix B: Coupling between twisted light and modes

We performed FDTD simulations using twisted light of different $l$ and polarization $\varepsilon_\sigma$. For each simulation, we confirmed if there is coupling between the excitation and the nanocavity modes. If there is significant excitation by the twisted light, we look at the rotation of the modes to deduce the traveling mode component that the light couples to. The results are summarized in Table 1.

Following from this, we deduce that the selection rules should arise from the coupling term between the twisted light and the nanocavity modes of the following form:

$$\int E^{\pm*}(\phi) E^{TL}(\phi) d\phi \propto \int_0^{2\pi} \exp\left[i\phi(-m+l+p_l)\right] d\phi \quad (9)$$

where on the right hand side $-m$ originates from the complex conjugate of the eigenmodes, $l$ being the topological charge and $p_l$ is the contribution from $\varepsilon_\sigma$. For non-zero coupling, this azimuthal integral term expressed on the right hand side of Eq. (9) needs to be nonzero and thus the integrand needs to be 1 by having the exponents summing up to zero. Therefore, the phases and the polarization of the relevant fields give a selection rule of

$$l + p_l = m \quad (10)$$

reflecting the conservation of angular momentum. We confirmed the validity of the selection rules by performing also simulations with other values of $l$. The results for $l = 3$ which could give non-zero coupling is listed also in the Table 1. The cases for $l = 0$ and 2 gives no coupling with the nanocavity mode as predicted by the selection rules. The non-coupling is most likely due to the antisymmetric nature of the quadrupole mode electric fields which results in the negligible overlap with excitation of $l = 0$ and 2 when integrated over the whole of the nanocavity.

| $l$ | $\epsilon_\sigma$ | $p_l$ | $l + p_l$ | Coupling | Mode rotation | Rotating mode, |
|---|---|---|---|---|---|---|
| +1 | RCP | +1 | +2 | Yes | Clockwise | +2 |
| +1 | LCP | -1 | 0 | No | - | - |
| +1 | Linear | +1, -1 | +2, -2 | Yes | Clockwise | +2 |
| -1 | RCP | +1 | 0 | No | - | - |
| -1 | LCP | -1 | -2 | Yes | Anti-clockwise | -2 |
| -1 | Linear | +1, -1 | +2, -2 | Yes | Anti-clockwise | -2 |
| +3 | RCP | +1 | +4 | No | - | - |
| +3 | LCP | -1 | +2 | Yes | Clockwise | +2 |
| +3 | Linear | +1, -1 | +2, +4 | Yes | Clockwise | +2 |
| -3 | RCP | +1 | -2 | Yes | Anti-clockwise | -2 |
| -3 | LCP | -1 | -4 | No | - | - |
| -3 | Linear | +1, -1 | -4, -2 | Yes | Anti-clockwise | -2 |

**Table 1. A summary of FDTD simulation results with twisted light excitation of different combinations of $l$ and $\epsilon_\sigma$, along with the observation of coupling.**

## Funding

This work was supported by JSPS KAKENHI Grant-in-Aid for Specially promoted Research (15H05700), Grant-in-Aid for Scientific Research (B) (17H02796) and Grant-in-Aid for Scientific Research on Innovative Area (15H05868).

## Acknowledgement

We thank K. Kuruma for his technical support and many helpful discussions.